\documentclass[12pt]{amsart}
\usepackage{amsfonts,amsmath,graphicx,subfigure,longtable,supertabular}
\usepackage{times}      
\usepackage{latexsym}   
\usepackage{amssymb}    
\usepackage{amsmath}    
\usepackage{amsthm}     
\usepackage{mathrsfs}   
\usepackage{euscript}   
\usepackage{epsfig}             
\usepackage[all]{xy}    
\begin{document}

\title{Topological structures in the equities market network}

\author{Gregory Leibon}
\address{Department of Mathematics, Dartmouth College, Hanover, NH 03755}
\email{gregory.leibon@dartmouth.edu}
\author{Scott D. Pauls}
\address{Department of Mathematics, Dartmouth College, Hanover, NH 03755}
\email{scott.pauls@dartmouth.edu}
\author{Daniel Rockmore}
\address{Department of Mathematics, Dartmouth College, Hanover, NH 03755}
\address{Department of Computer Science, Dartmouth College, Hanover, NH 03755}
\email{daniel.rockmore@dartmouth.edu}
\author{Robert Savell}
\address{Thayer School of Engineering, Dartmouth College, Hanover, NH 03755}
\email{scott.pauls@dartmouth.edu}

\begin{abstract}

We present a new method for articulating scale-dependent topological descriptions of the network structure inherent in many complex systems. The technique is based on
 ``Partition Decoupled Null Models,''  a new class of null models that  incorporate the interaction of clustered partitions into a random model and generalize
the Gaussian ensemble.   As an application we  analyze a correlation matrix derived
from four years of close prices of equities in the NYSE and
NASDAQ. In this example we expose (1)  a natural structure
composed of two interacting partitions of
the market that both agrees with and generalizes standard notions of scale
(eg., sector and industry) and (2) structure in the first partition that is a topological manifestation of a well-known pattern of capital flow called 
``sector rotation.''  Our approach gives rise to a natural form of multiresolution analysis of the underlying time series that naturally decomposes the basic data in terms of the effects of the different scales at which it clusters. The equities market is a prototypical complex
system and we expect that  our approach will be of use in understanding a broad class of complex systems in which correlation structures are resident.  
\end{abstract}


\maketitle
\section{Introduction} 

Complex systems often arise  as  a consequence of multilayered
interactions among a large population of diverse agents. For example, neural capabilities arise as a result of the interactions of clusters of neurons of similar function  \cite{neuron}.  Social networks often function as interacting hierarchies of sub-networks  \cite{watts1,watts2}, as do link networks for
webpages \cite{broder}.  The dynamics of the equities market is driven
by interactions among sectors,  which are in turn influenced by their
component industries as well as by the strategies of large
institutional traders \cite{gompers}.  The financial markets are of
particular interest for researchers in complex systems, as their intrinsically numerical nature provides a wealth of data for analysis and hypothesis testing.  The significant
complexity of the  web of interdependence in the markets has a natural and
informative mathematical formulation in terms of a network  encoding
the correlation structure of some underlying time series (e.g., price and
volume) that measures something of the state of the financial
instrument. Indeed, such correlation networks are an important class
of networks that fall naturally into the larger class of complex phenomena   in
which entities are related according to some measure of similarity in
a complex system.  

In this paper we present a new tool for decomposing these kinds of correlation networks --- the ``Partition Decoupling Method.'' It is an iterative method in which  spectral considerations  (i.e., eigenvalues of a relevant matrix) are used to identify significant clusters  via comparison to some relevant random model.  The effect of these clusters is  removed from the underlying data to
reveal a residual layer of interaction ready for another round of structural decomposition.   In this way,
the correlation network, as a summary of some complex system, provides
the nexus for an interesting symbiosis between the ideas of
multiscale data analysis and topological network analysis. The
iterated removal
of the ``cluster effect'' is akin (in spirit) to the well-known
``multiresolution analysis'' that accompanies wavelet decompositions
in signal and image processing (see  e.g., \cite{mr1,mr2}, as well as
\cite{mr3} which contains a more general view of multiresolution
analysis). It is likewise similar to  a factor or principal component analysis \cite{fa} that creates a
succession of approximations to a correlation matrix.  

Our approach produces a sequence of partitions of the network, each providing a topological
description of an aspect of the network structure.  This in turn gives rise to  natural hierarchical
decompositions of the underlying data stream. The hierarchical structure of the data is also manifested in a  multiscale 
structure in the correlation. The derived partitions
suggest a new class of null models introduced herein, the {\it
Partition Decoupled Null Model (PDNM)},  which incorporates the
different clusters into a random model.  A PDNM   is best understood
as a generalization of the widely used  Gaussian ensemble (GE) null
model in which there is a natural incorporation of the structural information associated with the partitions. The PDNM carries with it several
interacting partitions, each with its own geometric structure, making
it a more textured and potentially more powerful model for
comparison. We anticipate that the  Partition Decoupling Method (PDM)
will be of use in a variety of disciplines in which structure based on
similarity measures (e.g., correlation) is expected.

As an example, we
give a multipartition analysis of the correlation network of a
portion of the equities market.  A multiresolution, multipartition
decomposition of the equities market is plausible as the  global structure of these dynamic
entities is emergent from huge
numbers of local interactions resulting from several different factors
reflecting the dynamics of supply
and demand on various scales. 

Within  each partition, we expose a multiscale network in
which nodes at any given scale are aggregations of nodes at a finer
scale.  The nodes both echo and extend the usual notion of sector in
the market.  The articulation of topological structure yields our second main result --- the unsupervised discovery of non-trivial homology
(loops) in the network of clusters, reflecting the well-known
phenomenon of capital movement called ``sector rotation.''  

Ultimately, we reveal that the equities market is best described as a collection of  processes defined on interacting networks --- a  characterization shared by many diverse  complex systems. 
We demonstrate that by a careful decoupling of  network partitions,  we may peel apart the layers of  network structure to reveal subtle interdependencies among network components as well as residual network structures hitherto masked by more dominant network processes.

Our approach differs in some important ways from  previous
applications of clustering techniques to the hierarchical
decomposition of complex systems --- and particularly from previous efforts in the articulation of  ``market
topology'' as manifested in correlation networks derived from equities. The most important difference is that our model is not strictly hierarchical, but instead details the interaction  between a number of different partitions of the network.   Our method places no constraint on connectivity of the nodes, whereas purely hierarchical approaches constrain the complexity (in terms of connectivity) of the defined
nodes in some manner (e.g., as a tree \cite{mategna_hierarchical_1999} or with some fixed bound on
topological type \cite{tumminello_tool_}).  Furthermore, while our use of the
GE null model (see the Methodology section) as a means of identifying relevant
information in our clustering step is in the spirit of random matrix null models
\cite{plerou_etal,plerou_universal_1999,Stanleyetal},  our method
provides a more detailed description of a network by identifying relevant clusters across multiple interacting partitions.  We also note that, in contrast to our clustering method, cluster identification using localization of eigenvectors
(e.g. \cite{plerou_etal}) generally produces clusters which do not necessarily partition the entire set of equities (however, see \cite{KJ} for a single partition result).  

\section{Methodology}
There is a natural tension in the analysis of complex systems
between the desire to recognize the  complexity of a system in its entirety 
and the desire to conceptualize the system
processes in terms of  the interaction of a minimal collection of discrete 
components. Our methodology is
designed to preserve important aspects of  system complexity
typically lost in the application of dimension reduction techniques. The
Partition Decoupling Method  is a principled method for  generating   multipartition
descriptions of the system which effectively capture  both  the
dominant structures defining the system as well as second order structures which are often obscured by the actions of the dominant processes.  It involves combining two algorithms:  the Partition Scrubbing  Method and the Hierarchical Spectral Clustering Method.

\subsection{Partition Scrubbing  Method}

Beginning with a discrete sample space of nodes or entities indexed by $I =
\{1,\ldots, N\}$ with associated time series $D=
\{D(1),\ldots, D(N)\}$ each of length $T$, we
identify  a collection of {\em characteristic time series}  $\mathcal{V}$  which  capture some aspect of the
structure of the $D$ series.  Note that these need not be (and rarely are) independent. The idea is that each member of $\mathcal{V}$ summarizes some property of the time series in $D$ and  projection of $D$ onto the subspace
spanned by $\mathcal{V}$ yields a dimension reduced representation of
$D$. From this we then derive a decomposition of $D$ into two orthogonal components--
the projection of  $D$ onto $\mathcal{V}$ and a residual  component
$\mathcal{R}$.  The process may then be repeated on $\mathcal{R}$,
finding a new set of characteristic time series, projecting and
computing a second residual component.  Iteration may be continued
until ``failure,'' of which there are two types.   ``Partitioning Failure'' occurs when the correlation structure of the residual time series is indistinguishable from the Gaussian ensemble (note that depending on context, this could be replaced by other null models) and we cannot reliably find characteristic times series.  ``Projection Failure'' occurs when the characteristic time series are numerically linearly dependent.  In this case the projection on $\mathcal{V}$ does not have a unique representation in terms of the characteristic time series.  Our view is
that in each iteration, the removal of the effect of the characteristic
time series reveals residual structure that may have been masked by
the dominant behaviour.

To apply the Partition  Scrubbing Method,  we  start with a collection of normalized sequences  $D^{0}(i)$, and a choice  of clustering methods for each $0 \leq \alpha \leq m$  that,  given a collection   $D^{\alpha}(i)$, will produce a mapping $C^\alpha:I \rightarrow \{1,\dots,  | C^\alpha| \}$ where $  |C^\alpha|$ denotes  the number of clusters generated by the method (hence $C^\alpha$ is assumed to be onto).   We calculate the set of characteristic time series associated with the partition:
\[\mathcal{V}^{\alpha}_k=mean\{D^\alpha(i) | C^\alpha(i)=k\}\]
for $1 \le k \le |C^\alpha|$.  Then,
$\mathcal{V}^\alpha=\{\mathcal{V}^{\alpha}_1,\dots,\mathcal{V}^{\alpha}_{ | C^\alpha|}\}$.\footnote{Notice, this method can be generalized to any method of constructing the {\em characteristic time series}  $\mathcal{V}$.}  

Next, we scrub the partition to produce $D^{\alpha+1}(i)$ from  $D^{\alpha}(i)$.  To do so, we   decompose  $D^{\alpha}(i)$  into the sum of two components: the projection $\mathcal{F}^{{\alpha}}(i)$ associated with the clustering ${\mathcal{C}}^{\alpha}$  and a residual component $\mathcal{R}^{{\alpha}}(i)$, so that:
  \begin{equation}\label{cl-eq1}
D^\alpha = \mathcal{F}^{\alpha} + \mathcal{R}^{\alpha}
\end{equation}
where
\begin{equation}\label{cl-def}
\mathcal{F}^{\alpha}(i) = \Pi_\mathcal{V^{\alpha}}\left(D^\alpha(i)\right)=\sum_{k=1}^{{ |C^\alpha|}}
\tau_k^{\alpha}(i) \mathcal{V}_k^{\alpha}
\end{equation}
where $ \Pi_\mathcal{V^{\alpha}}$ is the projection onto $\mathcal{V}^{\alpha}$. 

We assume that
$\mathcal{R}^{\alpha}$-- the residual component of the time
series-- is independent\footnote{Here independence is meant in the statistical sense,  namely, that  they are not correlated. } 
of $\mathcal{V}^{\alpha}$.  Under these assumptions we can solve for the $\tau_k^{\alpha}(i)$ via some simple linear algebra.\footnote{We   take the inner product of both sides of \eqref{cl-eq1} with $\mathcal{V}_j^{\alpha}$ for all values of $j$ and solve for $\tau_k^{\alpha}(i)$. Equivalently: Let $A_{i,j}=corr(V_i,V_j)\cdot sd(V_j)$. Let $b(j)=corr(V_j,D^\alpha)\cdot sd(D^\alpha)$. Solve for $T=A^{-1}\cdot b$. Then $\tau_k^{\alpha}(i)=T(k,i)$.}  
We call $\tau_k^{\alpha}(i)$ the ``cluster  pressure
on node $i$'' (at iteration $\alpha$) .

 Given these values of $\tau$ we create a new collection of ``cleaned'' time series: 
\[D^{\alpha+1}= norm(\mathcal{R}^{\alpha})=norm(D^\alpha - \mathcal{F}^{\alpha}) \]
with $norm(\mathcal{R}^{\alpha})=\frac{\mathcal{R}^{\alpha}-m^{\alpha}}{\sigma^{\alpha}}$, where $m^{\alpha}$ and $\sigma^{\alpha}$  denote the mean and standard deviation of $\mathcal{R}^{\alpha}$ respectively.

Using this algorithm, each  series $D^{0}(i)$ can be reconstructed from the $D^{m+1}(i)$  from the $\sum_{\alpha=0}^m  |C^\alpha|$    characteristic time series in   $\{ \mathcal{V}^{\alpha}\}_{\alpha=0}^m$ and the   $\sum_{\alpha=0}^m  |C^\alpha|+2(m+1)$ parameters   $\{ m^{\alpha}(i),\sigma^{\alpha}(i),\{ \tau^{\alpha}_k(i)\}_{k=1}^{|C^\alpha|}   \}_{\alpha=0}^m$
corresponding to the entity. This is our ``multiresolution'' representation of the original time series data.\footnote{Note that Projection Failure occurs when the $\tau_k^{\alpha}$ are not uniquely determined (i.e., the matrix $A$ indicated in footnote 3  is not invertible).  We interpret  this as a loss of resolution in the data and/or a build-up of numerical error (and stop iterating if such a failure occurs).}

\subsection{Hierarchical Spectral Clustering Method}
To find the partitions  needed in the Partition Scrubbing Method, we use  an innovative hybrid technique, the Heirarchical Spectral Clustering Method.  This method is  a principled hierarchical clustering  of the correlation network, which proceeds by comparing the eigenvalues of the Laplacian of the correlation network to the eigenvalues of  a GE Null Model  associated with the network nodes. The method is suitable for networks in which effects of interest tend to result in stratification of network correlation strengths at particular scales.   Given a collection of time series indexed by $I$, the output of this method is $n$ levels of clusters of the nodes, each of which provides a partition of $I$.

At the core of the method is the dimension reduction via spectral clustering of the graph Laplacian \cite{Youngetal}
associated with the correlation matrix. (see the Supplemental Information for an overview of the method).    When presented with a correlation matrix for a sequence of time series, we identify the number of significant clusters and perform spectral clustering.  To pick the number of significant clusters, we are guided by the use of the 
GE null model as a means of determining at what point in the
spectrum of the Laplacian we are witnessing a manifestation of
 random effects.  $GE(n,m)$ models $n$ nodes with time series of length $m$, drawn from
 i.i.d. Gaussian random variables.  The choice of Gaussian random
 variables (as opposed to a different distribution) is motivated by
 our choice of application:  the total distribution from the
observed data for the equities network is close to Gaussian, with the
obligatory fat tails.\footnote {As a check, we performed our entire analysis with a bootstrap null model based on the observed data distribution but found no difference (in comparison with the use of a GE) in the results.  Thus, for ease of exposition and replication, we use the
Gaussian distribution as our base distribution. For other
applications, a different choice of distribution may be appropriate.}
We set the number of significant clusters equal to the number of nonzero eigenvalues of our correlation matrix  which fall below the minimum of the nonzero eigenvalues of the Laplacian of the correlation matrix associated with $GE(n,m)$ after simulating the distribution 100 times.
 
We call this first set of clusters  the first {\it level}.
 To form the remaining {\it levels}, we
repeat the following two steps until we reach a level with fewer than 2 clusters.  Given a level $j$,
\begin{enumerate}
\item[i.] Form a new correlation matrix $Corr(j)$ by computing the
  correlations between the mean time series of the clusters of the
  level $j$.
\item[ii.] Repeat the comparison to the GE null model and spectral clustering
  described above to find the $(j+1)^\text{st}$ level of clusters (i.e. these are
  clusters of clusters).
\end{enumerate}

 This step fails if at level one the 
comparison to the GE null model yields less than 2 significant
eigenvalues.  This we refer to as  a Partitioning Failure.   We call a level {\it nontrivial} if there are greater than 1  significant eigenvalues.

\subsection{Partition Decoupling Method  (PDM)}

The PDM consists of the iterative application of the Partition Scrubbing Method using the partitions  produced  by the Hierarchical Spectral Clustering Method.  As a first step,  we  normalize the series and we set $C^{0} \equiv 1$.   This is  akin to defining a partition
with a single characteristic time series $\mathcal{V}^0$ incorporating
all nodes.  (In our equities example, this corresponds to removing the global market effect by removing the overall daily mean, and  is similar to the normalization used in \cite{plerou_etal}.)  
Then  we proceed by using the  Hierarchical Spectral Clustering Method to form the partitions   needed by the Partition Scrubbing Method.   Notice, to run the Hierarchical Spectral Clustering Method requires choosing a level at each iteration, and we express these choices with the   {\it Partition Vector}  $\langle \ell_1,\dots,\ell_m \rangle$. A partition vector  uniquely determines the PDM's output: the characteristic time series   $\{ \mathcal{V}^{\alpha}_{\langle \ell_1,\dots,\ell_m \rangle}\}_{\alpha=0}^m$   
 and the constants 
 $\{ m^{\alpha}_{\langle \ell_1,\dots,\ell_m \rangle}(i),\sigma^{\alpha}_{\langle \ell_1,\dots,\ell_m \rangle}(i),\{ \tau^{\alpha}_{k}(i)  \}_{k=1}^{ | C^\alpha|}  \}_{\alpha=0}^m$
  for each entity.  Here  $C^{\alpha}_{\langle \ell_1,\dots,\ell_m \rangle}$ denotes the partition formed during the $\alpha$ iteration of the PDM.
   
Notice the  PDM  implicitly defines a restricted
class of  models via constraints on the covariance structures
associated with the traditional GE null model. We refer to these
associated null models as Partition Decoupled  Null Models (PDNM).  Given a
partition vector $\langle \ell_1,\dots, \ell_m \rangle$, we may construct
an associated PDNM by replacing the 
final $D^{m+1}$ with independent Gaussian random variables and inverting the Partition Scrubbing Method. 
Notice, if the decomposition terminates with a Partitioning Failure at  the $\alpha$ iteration,  then the $D^{\alpha+1}$ time series have a correlation structure that is  indistinguishable (in the above spectral sense) from the Gaussian ensemble, and this model duplicates the correlation network structure up to random effects. (For  decompositions that halt due to
a Projection Failure, the residual may still have significant
structure when compared to a GE null model, but we cannot reliably
compute the contributions of the clusters (i.e. the $\tau_k^{\alpha}(i)$)). 

This said, we view the importance  of the PDM not as providing a complete model of the system, but rather
  as providing a simple description of the complex system when the complexity is due interacting partitions.  For example, to capture the complexity of two interacting partitions  with $N$ and $M$ parts respectively  might   require  as many as $MN$   characteristic time series using a hierarchical method.  Using the PDM, one can potentially reduce this to $M+N$  characteristic time series, providing  much better dimension reduction.  Of course a real complex system may not be a simple interaction of $m$ partitions,  and different choices of partition vectors may produce different dimension reductions and reveal different structures.  We view the PDM as  a convenient way to produce a family of distinct dimension reductions, encoded in the tree of partition  vectors.  In  statistical learning situations,  this  family of dimension reductions  can prove to be a valuable asset during the model selection process.

\section{Decomposition of Equities Networks}

For our specific application to the equities market network, we begin with $N$ time series  of daily close prices and  create an initial collection of series $D^{0}$ which corresponds to the logarithmic return (logarithmic derivative or fractional change) of the closing price series for each equity.  That is,  given $P_t(i),$ the closing price on day $t$ of equity $i$ we  approximate  the logarithmic derivative of  $P_t(i)$ by:
\[D^0_t(i)= \frac{P_t(i)-P_{t-1}(i)}{P_{t-1}(i)} . \]

In this section, we  describe the results of applying the PDM  to the equities network determined by these series.  We demonstrate the ability of the PDM to expose network structures which elude typical clustering methodologies. 
In doing so, our results delineate a more general notion of market sectors than those typically acknowledged by the industry, in  that we expose both recognizable ``classical sectors'' as well as new natural hybrids. Additionally, the coarse scale analysis successfully exposes a non-trivial homological entity (a topological cycle) corresponding to the known phenomenology of capital flow referred to as ``sector rotation.''  

For this application, we obtained from the {\em Yahoo!} Finance
historical stock data server  daily close prices for 2547 stocks
currently listed on the NYSE and NASDAQ for a period spanning  1251
trading days over (roughly) four years  (in the time window from March 15,
2002 - Dec 29, 2006). We  began by removing any equity with more than
30\% missing data in that window, after which we were left with our 2547
equities over 1251 trading days.  In addition, we remove all extreme
events from the time series (20 \% or larger single day moves).  This cleaning was performed in part to avoided having to carefully compensate for the stock splits and reverse stock splits in our data.    
But we feel that this cleaning would  be appropriate even if we had cleaned out the splits via other methods. This is because the structure that underlies the market 
exists in (at least) two regimes --- extreme events and ``normal''
events --- articulated as two different network structures \cite{lo}. 
Since the extreme events were very sparse, the time  series correlations we used to explore the equities market by their nature are only capable of  illuminating the  ``normal'' network.

\subsection*{PDM Applied the the  Equity Market}
To demonstrate the method's superiority in exposing latent structure
in the network, we look at the results of the PDM  for two iterations.  This  resulted in four possible 
partition vectors with non-trivial levels, as schematically described  in figure  \ref{nullcomp}. 

 We found partitions of the following  sizes:   $\left| C^{1}_{\langle 1,* \rangle} \right| = 49$,  $\left| C^{1}_{\langle 2,* \rangle} \right| = 7$,  $\left| C^{2}_{\langle 1,1 \rangle} \right| = 62$, $\left| C^{2}_{\langle 1,2 \rangle} \right| = 10$, $\left| C^{2}_{\langle 2,1 \rangle} \right| = 52$, and $\left| C^{2}_{\langle 2,2 \rangle} \right| = 10$. 
Notice, the partition at the first iteration is independent of  future iterations, hence the $*$ can denotes any choice.
All of these partition vectors provide effective and distinct dimension reductions of the overall complex
system.  We now explore these examples with the goal of demonstrating that PDM results  in a collection of effective  dimension reductions and additionally uncovers information that is often obscured using traditional clustering decompositions.

For each partition, we use the industry sector labelings available from
{\em Yahoo!} Finance and NASDAQ/NYSE membership to examine the composition of clusters as
both a
validation of our clustering method and a tool which helps show when
partitions reveal new information.  We find that the majority (35 of
49) of the clusters of $C^{2}_{\langle 1,* \rangle}$ are predominantly identified by sector (in the sense that the majority of their nodes are from a given sector) and
most of the clusters are strongly identified with either the NASDAQ or
the NYSE (see Supplemental Information, Figure
\ref{sectorfig}).  Seven of the clusters without dominant
sectors have other obvious categorizations (e.g., a regional or
business commonality).  
 
\begin{figure*}

\includegraphics[scale=0.5]{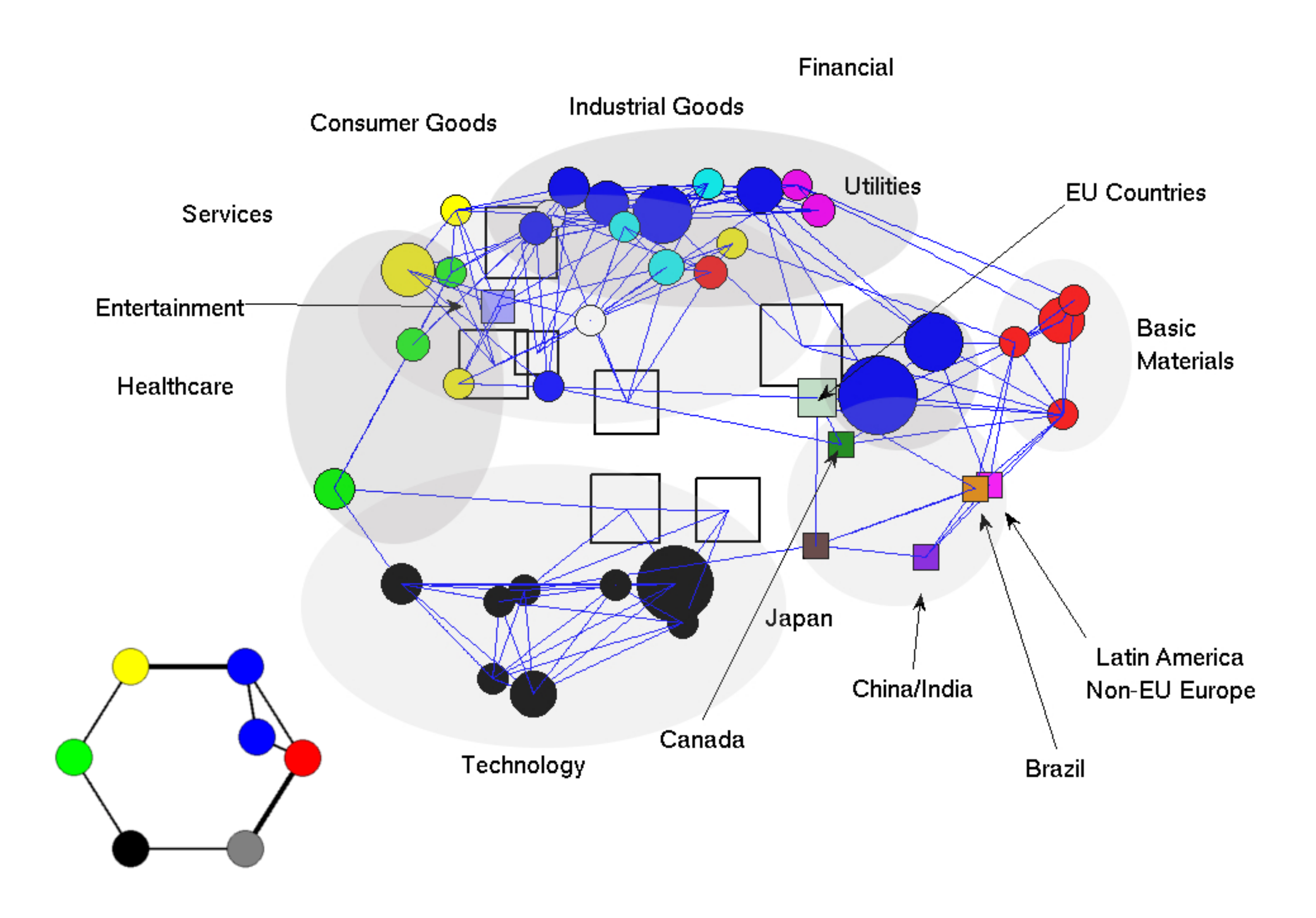}

\caption{Network structure after after applying PDM and dimension
  reduction.  The big graph is $C^{1}_{\langle 1,* \rangle} $ with distance  determined by the 
  correlation between the resulting characteristic time series.  Solid circular nodes are classified clusters with coloring  indicating the dominant sector or classification.  Unfilled square
  nodes are clusters without a dominant classification
  labeling.  Node size in all cases is proportional to cluster size.
  Connections (blue lines) are added when the Euclidean distance
  between two cluster centroids in the Euclidean embedding  is in the bottom 10\% of all such
  distances.  The grey regions identify clusters of clusters and are (basically) 
   $C^{1}_{\langle 2,* \rangle} $.  A  schematic drawing of the resulting network is in the lower left hand
  corner.  Nodes are labelled 1-7 counterclockwise beginning with the yellow node.\label{cyclefig}}
\end{figure*}

Clusters of partition $C^{1}_{\langle 2,* \rangle} $  were also classified generally by
sector.  Figure \ref{cyclefig} shows a representation of the network
resulting from the spectral clustering algorithm applied in our first iteration.  For visualization purposes, we have
used the centroids of the clusters in $C^{1}_{\langle 1,* \rangle} $
to represent the entire cluster and have used standard
multidimensional scaling (see e.g., \cite{kmeansref}) to reduce to a lower
dimension.  The grey regions in Figure
\ref{cyclefig} roughly reflect the clusters of $C^{1}_{\langle 2,* \rangle} $.  The inset graph
in the lower left hand corner shows only the $C^{1}_{\langle 2,* \rangle} $  clusters and is colored according to dominant sector. Tables \ref{T2} and \ref{T2.1}
provide a precise summary of the clustering data and classification.
Clusters of $C^{2}_{\langle 2,1 \rangle} $  predominantly admit natural
classification (30 out of 52 are classified by sector/industry and 5
more are classified by geography) while the opposite is true of clusters of
$C^{2}_{\langle 2,2 \rangle} $, where only 3 of 10 admit sector classification (witnessed in Figure \ref{fig3}).

The clusters of  $C^{2}_{\langle 2,2 \rangle} $ and  $C^{2}_{\langle 2,1 \rangle} $ 
provide new partitions of the network and reveal new, textured
information previously obscured by behavior of the dominant clusters
discovered in the first iteration.  While clusters of both $C^{1}_{\langle 1,* \rangle} $ and  $C^{2}_{\langle 2,1 \rangle} $  are classified by sector and have
significant membership overlap, the network configuration is substantially different from that shown in Figure \ref{cyclefig}.  This demonstrates that the clusters of   $C^{2}_{\langle 2,1 \rangle} $  correspond to a new subsidiary network structure,  revealed by exposing new strata of correlation strengths (of lower magnitude) previously masked by the dominant behavior of the  clusters in   $C^{1}_{\langle 2,1 \rangle} $.
 While the original clustering  of the nodes in   Technology  in $C^{1}_{\langle 2,1 \rangle} $ were positively correlated and tightly grouped,  the removal of  $C^{1}_{\langle 2,1 \rangle} $   via partition decoupling  exposes a new configuration for these entities in which there is clustering in similar groupings but with different internal relationships,  including negative correlations.  In Table  \ref{lay1-table},
we see the  classification of the clusters into which this technology cluster ($C^{1}_{\langle 2,1 \rangle} $) decomposed  in  $C^{2}_{\langle 2,1 \rangle} $.    It is evident from
this analysis that the partition decoupling has removed the major
effect of  $C^{1}_{\langle 2,1 \rangle} $, revealing lower order
effects as expected.  We hypothesize that these new partition layers may indicate
``second order''  trading strategies within these sectors.   We note
that  within  the other clusters of $C^{1}_{\langle 2,1 \rangle} $, similar
effects are found. 

The representation  of  $C^{2}_{\langle 2,2 \rangle} $  is
shown in Figure \ref{fig3}.  The three clusters classified by sector reflect reconfigurations of the sectorial
divisions given by  $C^{1}_{\langle 2,2 \rangle} $.  More interesting are the
unclassified clusters which reveal new cross-sector interactions.  For
example, the diamond shaped clusters  contain a mixture of multiple
sectors. The first is predominantly Consumer Goods, Industrial Goods
and Services, while the second is predominantly Financial, Healthcare,
Services and Technology. However, both clusters contain significant
commonalities.  In the first, the equities in the Service sector are
almost all related to the shipping industry, which obviously serves to
distribute Consumer and Industrial Goods.  In the second, the equities
in the Financial, Services and Technology sectors are related to
companies that either provide services or do business with healthcare
companies (e.g. health insurance companies, drug companies, management
services, healthcare based REITs, etc.).  Equities in both of these
clusters are drawn from a range of different clusters in $C^{1}_{\langle 2,2 \rangle} $,
showing that these two overlapping partitions are truly distinct, and once again demonstrating PDM's 
to remove higher order effects and reveal new structure.

\subsection*{Nontrivial homology - sector rotation}

The most significant geometric property of the hierarchical network
exposed in the first iteration is the existence of  a topological cycle (i.e., an example of nontrivial homology) 
reflective of the well-known phenomenon of  {\em sector rotation} ---  which forms the basis for predictive techniques in Intermarket Analysis  \cite{sectorrotation}.   Sector rotation refers to the typical pattern of  capital flow  from sector to sector over the course of a business cycle. Capital flow is echoed in our network structure via enhanced correlations among related equities, and the topological cycle corresponding to sector rotation  manifests itself as an emergent structure in the dense network of  near neighbor links.

\begin{figure}
\includegraphics{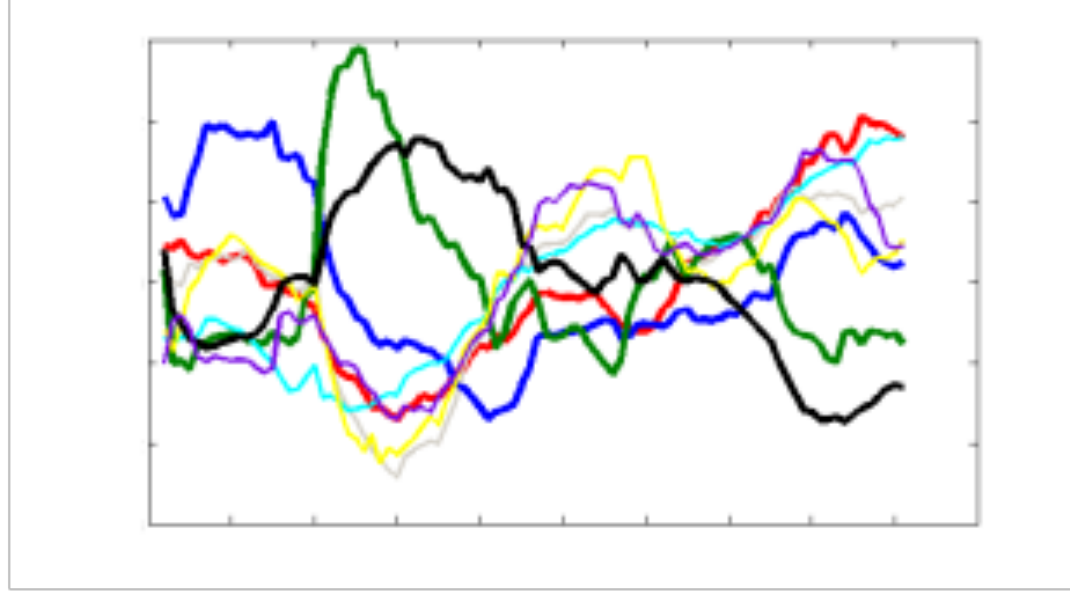}
\caption{Average $\tau$ by sector of time:  one year windows over ten
  years}\label{taurot}
\end{figure}

\begin{figure*}
\includegraphics[scale=0.6]{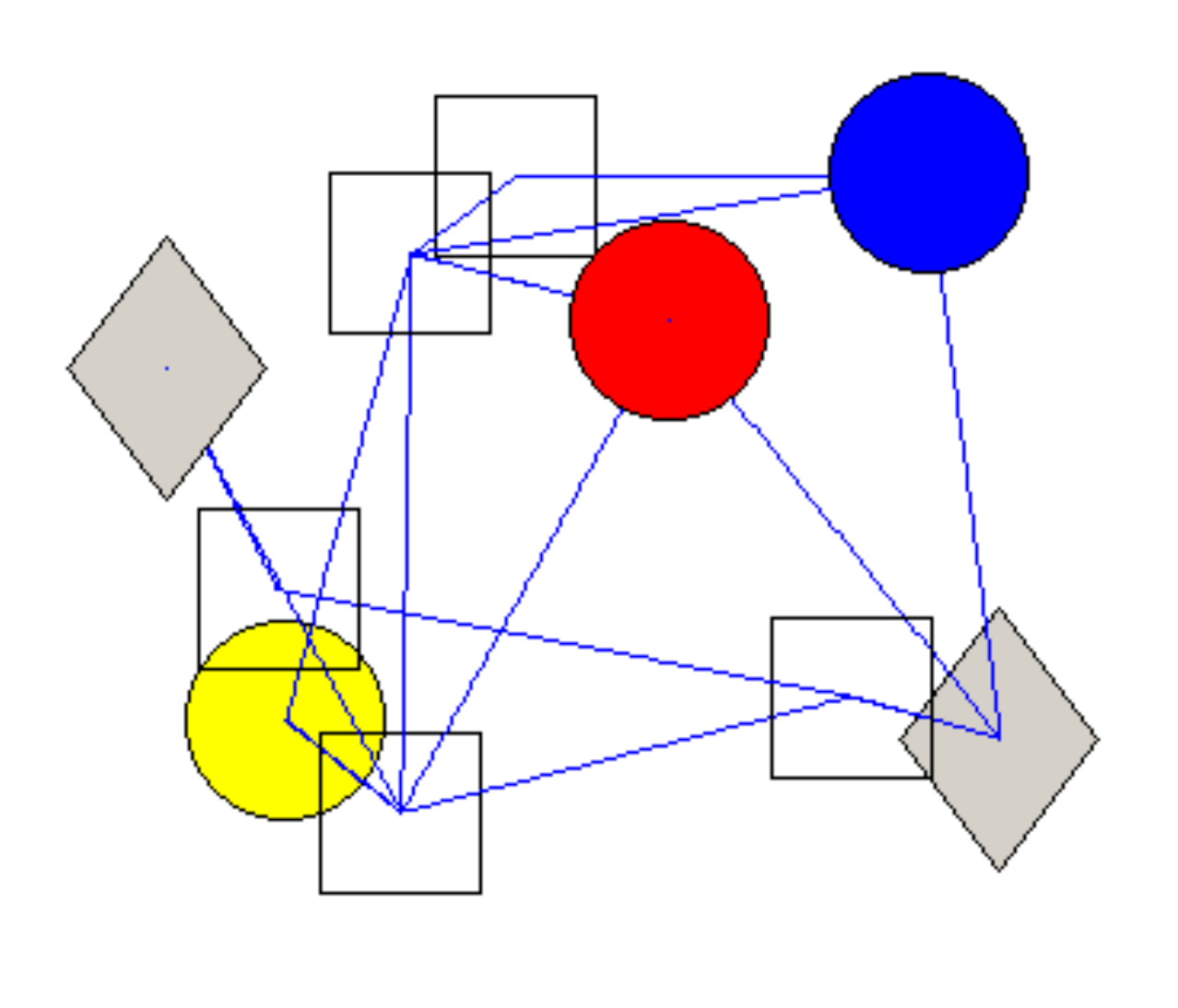}
\caption{The network with nodes determined by  $C^{2}_{\langle 2,2 \rangle} $ and 
with distance  determined by the    correlation between the resulting characteristic time series.
Three are identified by sector (red=Basic Materials,
  yellow=Services, blue=Financial) while the two diamond shaped
  clusters are classified by inter-sector commonalities as described
  in the text.} \label{fig3}
\end{figure*}

\begin{figure}
\includegraphics[scale=0.5]{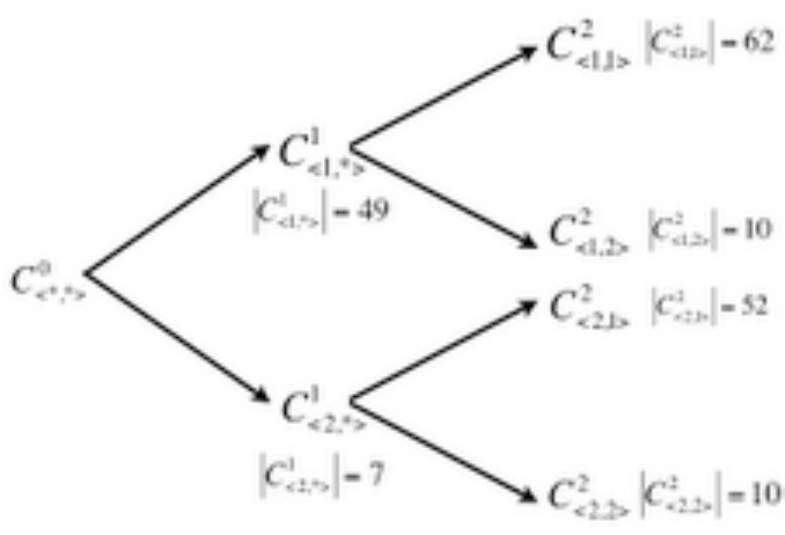}
\caption{A tree diagram showing the partitions involved when exploring two iteration 
of the Partition Decoupling Method with respect the equity market example. 
}\label{nullcomp}
\end{figure}

To support  the hypothesis that we are exposing sector rotation we compute the effect of the overall market pressure, $\tau$, for each equity in a moving one year window over ten years of
data.  As most of our clusters are sector dominated,  we compute,  as a proxy for
the aggregate pressure on the clusters, the mean $\tau$ for each
sector.\footnote{Recall that $\tau$ with respect to any subset (including the entire market) is the time series given by the average fractional change over the entire subset on each day.} In Figure \ref{taurot}, we plot the results over time after applying standard normalization. Both the periodicity of the sector effects and the relative phases of the sector waveforms  strongly  support the sector rotation  interpretation.  

The detection of a well-articulated  topology in this network is in a similar spirit to that of \cite{homology,homology2} where the computation of the homology of large datasets is used to topologically classify the data configuration.  In our case, the homology has a natural interpretation in terms of observed market behaviour.  

\section{Conclusion}
We present a new method for the  decomposition of complex
systems given a correlation network structure which yields scale-dependent geometric information --- which in turn provides a multiscale decomposition of the underlying data elements.  The PDM generalizes traditional multi-scalar clustering methods by exposing multiple partitions of clustered entities. 

Our multi-partition decomposition allows us to create a new class of null
models with which to study such systems:  the  Partition Decoupled Null
Model.  These null models mimic the observable clustering of the
network and thus provide a better
platform than the random matrix theory models from which to study the
behaviour of the network. 

As an example and application, we
analyze  a substantial portion of the US equities market, revealing several partitions that expose six different dimension reductions of the market network.  Labelling by traditional sector and
industry data validate  one aspect of the partitioning, as the finest partitions break down both
by traditional sector as well as other commonalities.  In addition to
validation of the technique (by recovering ``official''
classifications), the labelling provides evidence for our technique's
ability to extend traditional notions of a priori clusters in the
data. The partition decoupling reveals several instances of
cross-sectorial  components (with verifiable mixture classifications)
which tend to be obscured by the typical sectorial analysis. 

In the course of our decomposition we also identify an instance of
non-trivial homology: a cycle which corresponds to the well known
phenomena of sector rotation. This  ``sector rotation'' reflects the
movement of various sectors of the  equities market, which rise and fall in a predictable
cyclic manner as the economy moves through the stages of expansion and
contraction.  Our topological cycle in the correlation network
captures this phenomenon exactly as the order of the cyclic sector
rotation is reflected in the cyclic ordering of the cluster
components.  
 
In conclusion,  the PDM applied to the correlation network of the equities
market reveals both interesting known structure  and new structures
typically lost in common sectorial market decompositions. This
principled decomposition of the time series according to the structure
of the correlation network should prove useful for various forms of
risk management including portfolio construction. In addition, we
anticipate that other correlation networks produced by the actions of
diverse complex systems will also prove amenable to this approach.  

\section{Supplemental Information: Spectral Clustering}

We briefly outline the
procedure for spectral clustering used in \cite{Youngetal}.  Let $\rho$ be a correlation matrix for some set of nodes.
\begin{enumerate}
\item First construct the graph Laplacian associated to the
  correlation matrix, $\rho$:
\[L=I-D^{-\frac{1}{2}}\cdot \exp(-d^2)\cdot D^{-\frac{1}{2}}\]
where $d=\sin(\arccos(\rho)/2)$ is half the chordal spherical
distances associated to $\rho$, $D$ is the diagonal matrix with
entries given by the the row sums of $\exp(-d^2)$.
\item After computing the eigenspectrum $\{\lambda_i,v_i\}$ of $L$,
  choose the $k$ most relevant eigenvector/eigenvalue pairs.  Create
  the matrix $V$ with columns given by the selected $v_i$.
\item Normalize each of the rows of $V$ to have unit length.  
\item  Perform $k$-means on the data points using the rows of $V$ as the
  coordinates in Euclidean space of each node.  
\end{enumerate}

\begin{table}
\caption{Classification of clusters from Figure \ref{fig3}}\label{lay1-table}
  \begin{tabular}{cc}
\hline
{Number }&{Classification}\\
\hline

{ Technology Cluster}\\\hline
1 & \normalsize  Application/Business Software\\
2 &  \normalsize Semiconductor Equipment/Materials \\
3 &  \normalsize Semiconductor manufacture \\
4 & \normalsize  Network and Communication Devices\\
5 &  \normalsize China/India based internet/\\
 & \normalsize wireless/online community companies\\\hline
 \end{tabular}

\end{table} 



\begin{figure*}
\includegraphics[width=6in,height=7in]{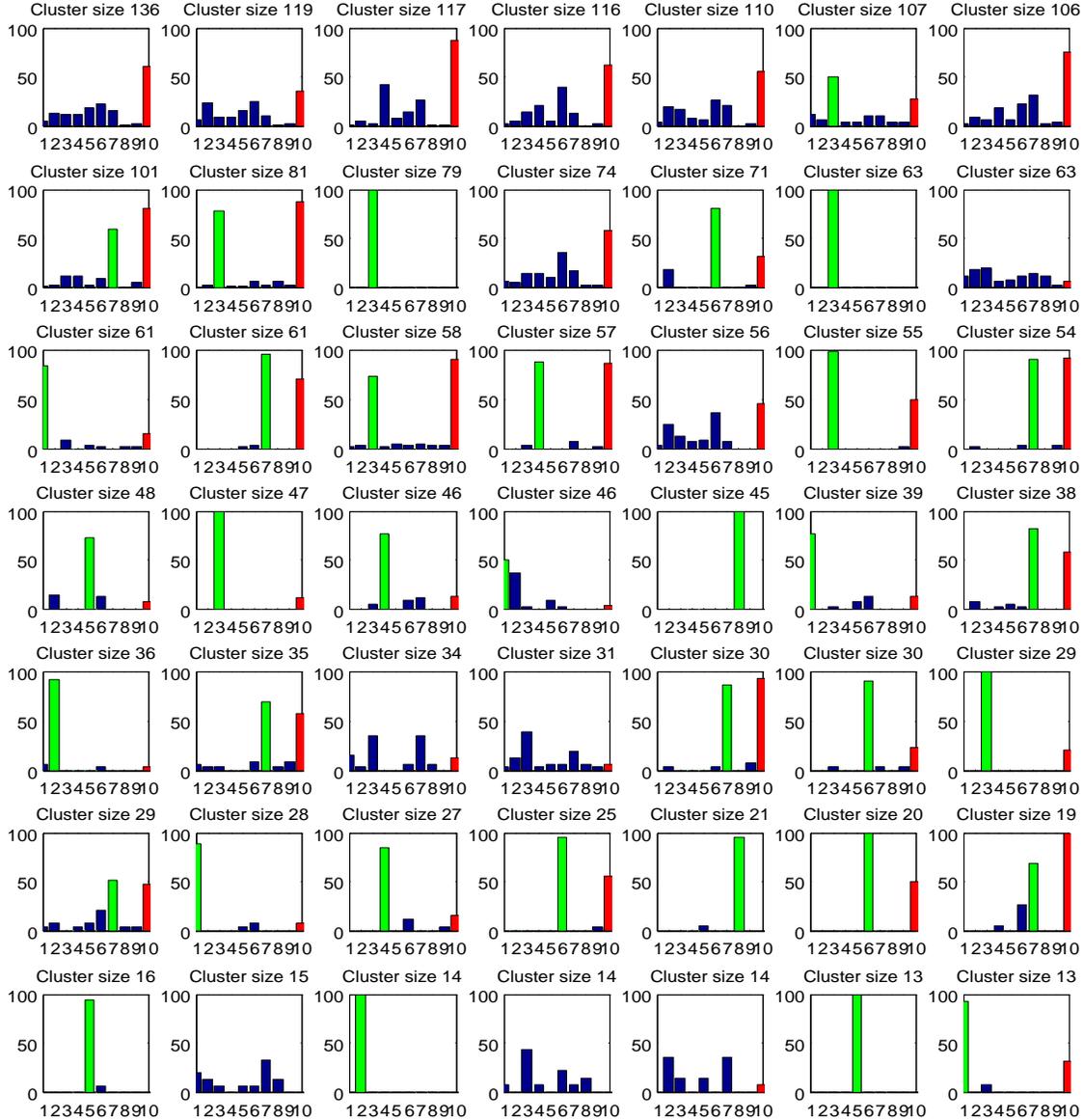}
\caption{Each figure contains a histogram of the sector and index
  breakdown of the clusters in $C^{1}_{\langle 1,* \rangle}$.  Columns 1-8 represent {\em
    Yahoo!} Finance sector labels. The ninth column is reserved
  for equities that had no sector labeling.  Height of each column
is the percentage of the total cluster falling in that category (the
cluster size is listed above each histogram).  A cluster is defined to
be sector dominated if one sector comprises more than 50\% of the
cluster.  In that case, the sector is denoted in green.   The tenth column, in red,
gives the percentage of the cluster that is listed on the NASDAQ.}\label{sectorfig}
\end{figure*} 
\begin{table}
\caption{Classification of clusters of $C^{1}_{\langle 1,* \rangle} $  and
 $C^{1}_{\langle 2,* \rangle} $ }\label{T2}
\begin{tabular*}{\hsize}{@{\extracolsep{\fill}}ccc}
\hline {\bf Cluster}$^*$ & {\bf Sector}$^{\dagger}$ & {\bf Classification}\\
(1,1)&N &   None \\  
(2,6)&N &   None \\  
(3,2)&N &   None \\  
(4,1)&N &   None \\  
(5,1)&N &   None \\  
(6,7)&F &  Closed End Funds \\  
(7,3)&N &   None \\  
(8,3)&T &   IT Products/Services \\  
(9,6)&F &   Regional Banking, S\&Ls \\  
(10,7)&F &   Closed-End Funds, Debt  \\  
(11,6)&N &   None \\  
(12,1)&S &   Strip Mall Stores \\  
(13,7)&F &   REITs \\  
(14,4)&N &   EU countries \\  
(15,5)&B &   Oil ans Gas \\  
(16,3)&T &   Semiconductors,\\ 
&&   Electronics \\  
(17,6)&F &   Regional Banking, S\&Ls \\  
(18,2)&H &   Biotechnology \\  
(19,1)&N &   Entertainment/Leisure \\  
(20,7)&F &   Regional Banking \\  
(21,3)&T &   Software \\  
(22,1)&I &   Construction \\  
(23,7)&F &   Insurance \\  
(24,2)&H &   Drugs/Medical Supplies \\ 
(25,1)&B &   Chemicals\\  \hline
\end{tabular*}
\noindent
${}^*$   The cluster is recorded as $(a,b)$ where 
 $a$ is the $C^{1}_{\langle 1,* \rangle}$ label and $b$ is the $C^{1}_{\langle 2,* \rangle}$ label.  
    \\${}^\dagger$ Sectors are identified via {\it Yahoo! Finance} labels as B
(Basic Materials), C (Consumer Goods), F (Financial), H (Healthcare),
I (Industrial Goods), N (None), S (Services), T (Technology), and U (Utilities)
\end{table}
 \begin{table}

\caption{Classification of clusters of $C^{1}_{\langle 1,* \rangle} $  and
 $C^{1}_{\langle 2,* \rangle} $ (continued) }\label{T2.1}

\begin{tabular*}{\hsize}{@{\extracolsep{\fill}}ccc}
\hline {\bf Cluster}$^*$ & {\bf Sector}$^\dagger$ & {\bf Classification}\\\hline
(26,7)&U &   Electric\\
(27,5)&B &   Industrial Metals \\ 
(28,3)&T &   Scientific/Technical\\
&&   Instruments \\  
(29,1)&C &   Grocery Store Items  \\  
(30,3)&T &   Communication \\  
(31,4)&N &   China and India \\  
(32,4)&N &   Latin America, Non-EU \\
&&  European countries \\  
(33,3)&T &   Computer components \\  
(34,1)&S &   Media Companies \\  
(35,7)&F &   Brokerages, Asset/\\
&&  Credit Management \\ 
(36,3)&T &   None \\  
(37,5)&B &   Oil and Gas Drilling\\  
(38,2)&H &   Health Care Plans \\  
(39,1)&S &   Shipping - Air and Rail \\  
(40,7)&U &   Gas \\  
(41,1)&S &   Restaurants \\  
(42,3)&T &   Internet Services \\  
(43,1)&I &   Aerospace \\
&&  Products/Services \\  
(44,4)&N &   Brazil \\  
(45,1)&C &   Auto Parts/\\
&&  Manufacture \\  
(46,7)&N &   Canada \\  
(47,4)&N &   Japan \\  
(48,1)&I &   Res. Construction \\  
(49,5)&B &   Gold Industries \\ \hline
\end{tabular*}
\noindent
${}^*$   The cluster is recorded as $(a,b)$ where 
 $a$ is the $C^{1}_{\langle 1,* \rangle}$ label and $b$ is the $C^{1}_{\langle 2,* \rangle}$ label.  
    \\${}^\dagger$ Sectors are identified via {\it Yahoo! Finance} labels as B
(Basic Materials), C (Consumer Goods), F (Financial), H (Healthcare),
I (Industrial Goods), N (None), S (Services), T (Technology), and U (Utilities)
\end{table}

\end{document}